\documentclass[manuscript]{acmart}
\usepackage{geometry}
\usepackage[most]{tcolorbox}
\usepackage{algorithm}
\usepackage{algpseudocode}
\AtBeginDocument{%
  }

\acmJournal{IMWUT}
\acmVolume{0}
\acmNumber{0}

\setcopyright{acmlicensed}
\copyrightyear{2018}
\acmYear{2018}
\acmDOI{XXXXXXX.XXXXXXX}

\usepackage{algorithm}
\usepackage{algpseudocode}

\begin{document}

\title[sMoRe]{sMoRe: Enhancing Object Manipulation and Organization in Mixed Reality Spaces with LLMs and Generative AI}

\author{Yunhao Xing}
\email{xyunhao@nd.edu}
\affiliation{%
  \institution{University of Notre Dame}
  \city{Notre Dame}
  \state{Indiana}
  \country{USA}
  \postcode{46556}
}

\author{Que Liu}
\affiliation{%
  \institution{University of Notre Dame}
  \city{Notre Dame}
  \state{Indiana}
  \country{USA}
  \postcode{46556}
}

\author{Jingwu Wang}
\affiliation{%
  \institution{University of Notre Dame}
  \city{Notre Dame}
  \state{Indiana}
  \country{USA}
  \postcode{46556}
}

\author{Diego Gomez-Zara}
\email{dgomezara@nd.edu}
\affiliation{%
  \institution{University of Notre Dame}
  \city{Notre Dame}
  \state{Indiana}
  \country{USA}
  \postcode{46556}
}
\additionalaffiliation{%
  \institution{Pontificia Universidad Cat\'olica de Chile}
  \department{Facultad de Comunicaciones}
  \city{Santiago}
  \country{Chile}
}
\renewcommand{\shortauthors}{Xing et al.}

\begin{abstract}
In mixed reality (MR) environments, understanding space and creating virtual objects is crucial to providing an intuitive and rich user experience. This paper introduces sMoRe (Spatial Mapping and Object Rendering Environment), an MR application that combines Generative AI (GenAI) with large language models (LLMs) to assist users in creating, placing, and managing virtual objects within physical spaces. sMoRe allows users to use voice or typed text commands to create and place virtual objects using GenAI while specifying spatial constraints. The system leverages LLMs to interpret users’ commands, analyze the current scene, and identify optimal locations. Additionally, sMoRe integrates text-to-3D generative AI to dynamically create 3D objects based on users’ descriptions. Our user study demonstrates the effectiveness of sMoRe in enhancing user comprehension, interaction, and organization of the MR environment.
\end{abstract}




\received{XXX}
\received[revised]{XXX}
\received[accepted]{XXX}

\begin{teaserfigure}
  \includegraphics[width=\textwidth]{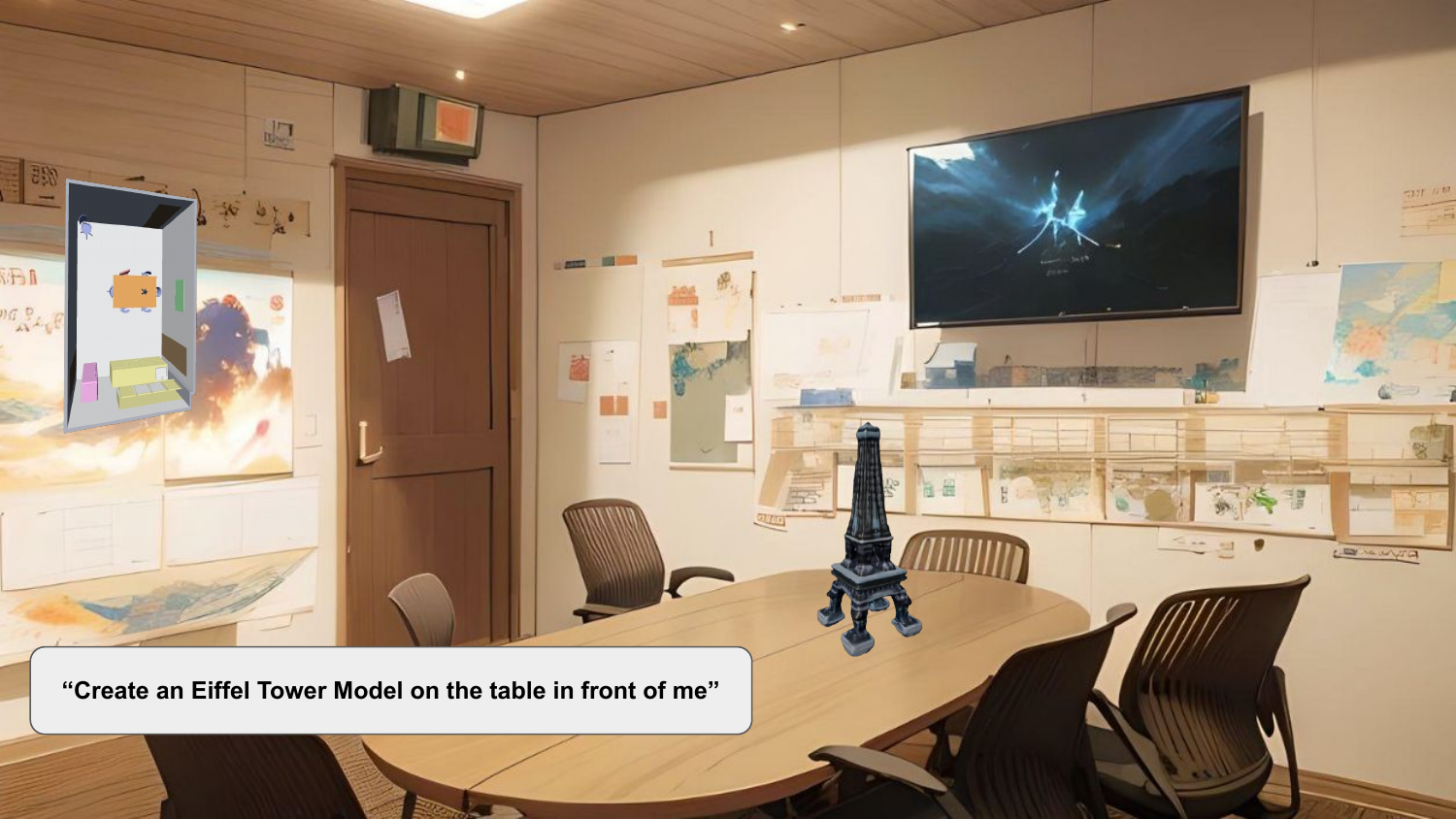}
  \caption{Illustration of the sMoRe system in action. Users provide voice or typed text commands, such as "Create an Eiffel Tower Model on the table in front of me," which sMoRe interprets to generate and accurately place virtual objects within their physical environment. The system also features a 2.5D layout map projected on the wall, showing simplified representations of all virtual objects in their relative positions.}
  \label{fig:teaser}
\end{teaserfigure}

\maketitle

\section{Introduction}
\label{introduction}
By blending virtual and physical worlds, mixed reality (MR) technologies have revolutionized how users interact with 3D virtual content. MR has emerged as a powerful tool in fields--ranging from education and entertainment to design and industrial applications--to enhance users' perception of spatial relationships and allow for dynamic, interactive experiences \cite{hololearn, aguayo2023using, gattullo2022design, tang2020evaluating, wu2023mr}. 

While generating objects in MR has been widely explored \cite{roberts2022stepspromptbasedcreationvirtual, RoomDreaming, HelpViz, NeuralCanvas, LLMR, Chamola}, the intuitive creation and placement of virtual objects within physical spaces remains a central challenge. Most systems do not provide a natural or intuitive way to create objects directly within the virtual space. Typically, creating digital objects in MR environments comes from pre-loaded libraries of assets limiting users to pre-designed options rather than allowing for spontaneous or creative object generation. Moreover, once these objects are created in the environment, their placement and manipulation often require the use of joysticks or controllers, which can be imprecise and not intuitive \cite{Yoffe2023,lee2024mixed}. Addressing these limitations is crucial for creating more fluid and engaging MR experiences. Thus, MR users need more practical interfaces to engage with the blended environment more naturally, allowing them to employ complex controls and enhancing their ability to organize and manipulate virtual objects. 

To address this challenge, we present ``sMoRe'' (Spatial Mapping and Object Rendering Environment), an MR application designed to empower users in creating, placing, and managing virtual objects through the integration of Generative AI (GenAI) and Large Language Models (LLMs). By combining the advancements in natural language processing and text-to-3D generation, sMoRe allows users to create and allocate virtual objects in MR environments using simple voice or text-based commands, reducing the cognitive and physical effort required to manipulate the virtual space significantly. 

This system takes advantage of the capabilities of LLMs to interpret user commands in natural language, considering the spatial constraints and context to optimize the placement of virtual objects. By incorporating text-to-3D generative AI, sMoRe offers the ability to dynamically generate custom 3D objects based on users' descriptions without needing specialized knowledge of 3D modeling or environmental design. Moreover, the system allows users to verbalize where they want to situate the new digital objects by analyzing their environment's layout and the relationships of existing physical objects. For example, users can say ``Create an Eiffel Tower Model on the table in front of me,'' and the system will understand that the virtual object should be placed on the physical table in front of the user (Figure \ref{fig:teaser}). 

In this paper, we describe the design and implementation of sMoRe, as well as the results of a user study that evaluates its effectiveness. Our findings indicate that sMoRe improves users' experiences in MR spaces by fostering a more organized and interactive interface between virtual and physical objects. As such, this paper highlights the potential of integrating GenAI and LLMs to create intuitive, user-friendly MR systems.

Our contributions are three-fold: 
\begin{itemize}
    \item We propose a novel framework that combines LLM-driven interactions with real-time generative AI capabilities to create and locate virtual objects in MR environments.
    \item We introduce intuitive and simple interaction methods to enhance the user experience in the MR environment, such as the 2.5D layout, voice commands, and hand/controller interactions.
    \item We conduct a comprehensive user study to validate the usability and effectiveness of this system and approach.
\end{itemize}
\section{Related Work}
\label{Related_Work}
We situate our work around prior studies examining (a) creating objects in MR, (b) interactive MR environments, and (c) using LLMs in MR.

\subsection{Creating Objects in MR}
The interaction with MR environments often involves creating and placing various virtual 3D objects. Creating objects in MR is usually achieved using user-generated content, generative AI, or fetching digital assets from online libraries \cite{roberts2022stepspromptbasedcreationvirtual, roberts2022surreal}. Examples are systems using \textit{Sketchfab}\footnote{\url{<https://sketchfab.com/>}} to identify virtual objects and fit them into the user's physical environment. Torre et al. \cite{LLMR} used DALL-E 2 to create a target image given the provided description, and then used \textit{CLIP} \cite{radford2021learningtransferablevisualmodels} to fetch and generate embedding of potential matches from \textit{Sketchfab}. In this way, the system could find the best digital object closest to the object-prompt sequentially based on language similarity and visual similarity \cite{LLMR}. Additionally, Rahaman et al. \cite{PhotoTo3D} introduced a methodology to produce 3D digital assets from photographs and later deployed it in an MR environment with inter-activity in supporting cultural heritage visualization and learning. 

Expanding on these approaches, Poole et al.'s \textit{DreamFusion} \cite{poole2022dreamfusiontextto3dusing2d} showcased the potential of utilizing a pre-trained 2D text-to-image diffusion model to generate diverse and intricate 3D objects. Lin et al.'s \textit{Magic3D} \cite{lin2023Magic3D} further improved \textit{DreamFusion} by implementing a two-stage optimization framework, achieving higher resolution and faster processing speeds. Meanwhile, Karnewar et al.'s \textit{Holodiffusion} \cite{Karnewar2023Holodifussion} leveraged diffusion models for scalable 3D generative modeling trained with only 2D images. Lastly, Gao et al. \cite{gao2022get3d} introduced \textit{GET3D}, a generative model that produces textured meshes directly usable by 3D rendering engines, facilitating immediate integration into downstream applications such as \textit{Unity3D} and \textit{Blender}.

The interplay between GenAI for 3D assets and extended reality (XR) has become a promising field of study. Chamola et al. \cite{Chamola} further explored the impact of generative AI on the immersive virtual world by delving into four domains: text generation, image generation, video generation, and object generation. Roberts et al.\cite{roberts2022stepspromptbasedcreationvirtual} advanced virtual object generation in MR by investigating voice prompts as a method to ``speak the world into existence'' and transform the objects of a virtual world. Lastly, Torre et al. \cite{LLMR} extended beyond just modifying visual appearances by incorporating interactions and behaviors into the generated 3D content. 

Moreover, recent research shows a growing trend in employing GenAI models for 3D asset generation. These models are trained on extensive datasets and can produce content across various mediums, such as images, text, audio, and video \cite{DivergentThinking}. Consequently, extensive research has investigated the influence of GenAI systems on user experiences in both creative and practical tasks, including interior design exploration, mobile tutorial generation, and scenic prototyping. For example, Wang et al. \cite{RoomDreaming} developed \textit{RoomDreaming}, a GenAI-based website that enables homeowners and designers to rapidly iterate through a wide range of photo-realistic design alternatives based on room layouts and preferences, thus enhancing the efficiency and satisfaction of preliminary interior design exploration. Zhong et al. \cite{HelpViz} introduced \textit{HelpViz}, a tool that converts text-based mobile phone instructions into visual tutorials by parsing actions from text instructions, executing them in Android emulators, and integrating graphical assets into text tutorials. With features like user progress tracking and next-step highlighting, \textit{HelpViz} improves tutorial execution robustness and user preference. Additionally, to replace vague sketching and time-intensive 3D modeling, Shen et al. \cite{NeuralCanvas} introduced \textit{Neural Canvas}, a 3D scenic design prototyping platform that seamlessly integrates 2D sketching with generative AI models. Designers begin with basic sketches and the AI models then enrich the sketches with intricate details. This approach significantly streamlines the scenic design prototyping process, enabling users to rapidly explore visual concepts and produce more detailed and attractive 3D scenic designs.

Despite the advanced work about creating objects in MR, the real-time generation of 3D objects directly within MR environments has not been explored. Our work tackles this gap by helping users to create and manipulate virtual objects in MR environments.

\subsection{Interactive MR Environments}
Interaction techniques for 3D objects in MR environments, including hand gestures \cite{ManipulationSurvey, handGestureDef}, controllers, and voice commands, have evolved considerably over the years. Several approaches based on multiple techniques allow users to interact with virtual objects as though they were real, leading to their widespread adoption due to their simplicity, intuitiveness, and immersive nature \cite{objectSelectionDef, Gaze-supportManipulation, Gogo, gestureKnitter, tangibleHandInteraction, virtualHand}. 

The engagement with MR can be further enhanced by improving the accuracy of object manipulation and expanding the genres of interaction. Traditional methods for object selection often rely on proximity-based heuristics, which infer the closest object as the intended target. This approach can lead to errors in densely populated areas \cite{predictiveSelection}. Recent efforts have explored various approaches for more accurate intention evaluation. For instance, Asano et al. designed algorithms that predict endpoints based on observed trajectory fractions \cite{predictiveAlgorithm}. Clarence et al. utilized users' reach trajectories to train neural networks for personalized predictions \cite{neutralNetworks}. Moon et al. developed an inference model based on bio-mechanical simulation, which reduces reliance on extensive human datasets and enhances task transfer capabilities \cite{bioMechanical}. Furthermore, there is a need for more refined object manipulation techniques. Current virtual hand implementations typically manipulate all six degrees of freedom (DoF) simultaneously, making it challenging to adjust only one dimension while keeping others fixed \cite{degreesOfFreedom}. In contrast, raycasting does not support rotation about axes other than the ray itself, nor does it support scaling \cite{RaycastingManipulation, virtualPointing}. Additionally, existing techniques primarily focus on manipulating the fundamental spatial properties of objects. Effective methods for refining finer details, such as color, texture, non-linear movements, or behavioral attributes, remain limited.

Instead of using hand inputs, voice commands are becoming a more and more trending way of interaction due to their simplicity. Specifically, intelligent virtual agents are built to perform tasks on behalf of user's questions or commands \cite{virtualAgents}. Users' voices are first converted to text using technologies like Microsoft Azure, and then the virtual assistant will execute the command according to the transcripts. The behavior is typically well-defined and allows triggering actions based on a set of conditions \cite{1996software}. As Ishigaki et al. \cite{ishigaki2023voice} explored in their work, the voice command with a virtual assistant avatar in MR telepresence can improve the interaction, which is more flexible and faster without physical constraints. 

Given the benefits of voice commands, our work focuses on improving the interaction between the MR environment and users using voice commands.

\subsection{LLM in MR}
Interacting with and manipulating virtual objects within MR environments can be complex, especially when user input is limited to traditional controllers or gestures. Recent research has focused on utilizing LLMs to bridge this gap, enhancing the user experience and enabling more intuitive and intelligent interactions within MR environments.

One area of focus has been implementing LLMs for the interpretation and representation of 3D objects in MR spaces. Luo et al. \cite{3DCaptioning} introduced ``Cap3D,'' a system designed to generate detailed captions for 3D objects. By refining and summarizing preliminary captions from 2D views with advanced models such as BLIP2, CLIP, and GPT-4, Cap3D allows users to gain a richer understanding of the virtual objects they encounter, enhancing accessibility and comprehension within the MR environment. This approach helps users engage with 3D objects more naturally, moving beyond static 2D representations.

To further advance spatial understanding in MR using LLMs, Xu et al. \cite{pointLLM} proposed PointLLM, which incorporates a point cloud encoder to perform viewpoint-independent analysis of 3D environments. This development allows LLMs to process and interpret spatial data more effectively, offering users a more robust understanding of the virtual world regardless of their perspective. Beyond object understanding, LLMs have also been applied to enhance interactions between users and virtual environments. Volum et al. \cite{NPC} and Wan et al. \cite{10.1145/3613905.3651026} demonstrated that LLMs can be used to guide the behavior of non-playable characters (NPCs) within virtual environments, enabling more realistic and responsive interactions in a more contextually aware manner. In most recent research, ``SituationAdapt'' \cite{li2024situationadaptcontextualuioptimization} was introduced as a system to adjust user interfaces in MR based on environmental and social factors, utilizing the contextual reasoning awareness of a Visual-and-Language model.

In terms of direct user interaction with virtual spaces, Manesh et al. \cite{10.1145/3643834.3661547} explored how voice prompts can be used to modify virtual environments, focusing on proxemic variables such as the spatial relationships between objects and users. Their system allows users to adjust and manipulate the virtual space using natural language commands, significantly reducing the need for technical expertise and making the environment more accessible to a wider range of users.

Programming and behavior generation within VR environments can often be challenging for users who lack coding experience. Addressing this challenge, ``DreamCodeVR'' \cite{10494096} introduced a system that simplifies the process of behavior generation in virtual worlds, enabling users to generate complex behaviors without needing in-depth coding knowledge.

Finally, 3D objects are fundamental to interaction within MR environments, and understanding these objects is crucial for effective user interaction. Lee et al. \cite{10392933} proposed a knowledge generation pipeline that assists users in comprehending 3D object information more effectively. Additionally, recent studies have also explored the role of LLMs as conversational agents within MR, VR, and AR environments. For example, articles by \cite{10.1145/3626705.3631799, misc} leveraged the natural language understanding and contextual reasoning capabilities of LLMs to assist users with simple tasks, demonstrating the potential for more intuitive and fluid user interactions within digital spaces. By employing LLMs as conversational agents, these systems create a more natural interface for users, allowing them to interact with virtual environments through dialogue rather than complex commands or gestures.

Building upon this research, our system aims to integrate the capabilities of GenAI and LLMs to enhance the users' experience within MR environments.

\section{sMoRe: Spatial Mapping and Object Rendering Environment}
\label{system}

\subsection{Design Goals}
sMoRe aims to expand the use of generative AI and LLMs in MR environments. The goal is to help users create, manipulate, and organize the blend of virtual objects within their physical spaces. To achieve this goal, we identified three design goals that will guide our proposed system's development.

\paragraph{DG1: Enhancing the user's ability to perceive, understand, and operate both physical and virtual spaces.} Understanding the spatial layout, object placement, and interactions between physical and virtual elements is essential for effective interaction in an MR environment. The system should promote spatial awareness by providing clear and intuitive visual and audio aids that help users comprehend both the physical and virtual components of the space.

\paragraph{DG2: Ensuring precise and fast placement of virtual objects within the MR space.} When creating virtual objects, users should be able to allocate them where they expect to see them, respecting physical space constraints. Thus, the system needs to accurately interpret spatial relationships from the user's instructions, process contextual information about the MR environment, and ensure that virtual objects are positioned with high precision. Additionally, the system must operate efficiently, minimizing delays between users' input and object placement to maintain a responsive user experience.

\paragraph{DG3: Providing intuitive spatial manipulation and organization.} Users should easily understand how to manage the virtual elements within the MR world without challenging learning curves. To achieve this goal, the system should offer intuitive gestures, voice commands, or natural interactions to generate, modify, and organize virtual objects in MR.

\subsection{Example Scenario}
We provide an example that illustrates the practical functionality of the sMoRe system. Maya, a busy professional, often forgets small but important tasks throughout the day, such as picking up laundry, watering the plants, or taking medicine. To help her manage her tasks, Maya uses sMoRe to set virtual visual reminders in specific locations around her home, ensuring that reminders are visible and placed exactly where she notices them. 

After the MR head-mounted device (HMD) scans her apartment, Maya starts using sMoRe. She sees the physical space and notices that her car keys are on one of the cabinets. To remember the location of the keys, she gives the following voice instruction to sMoRe: \textit{``Put a large car key on the cabinet in front of me.''} Upon finishing the sentence, she sees the voice instruction displayed as text in front of her. This immediate visual feedback reassures her that the system has understood her request, or whether it must be clarified. After a few seconds, the system shows Maya a floating pop-up window in the middle of the room and displays four different designs of virtual car keys from which to choose. These keys were generated using GenAI from the user's voice instruction. Maya selects the version that resembles her car key the most. She can use the raycast from her controller or her hands to select the virtual car key. The pop-up window disappears, and she sees a blue highlighted area appear on the surface of the cabinet, indicating where the car key reminder will be placed. After a few seconds, the virtual car key that she selected appears in the blue area, seamlessly integrated into her physical space as a virtual reminder. Then, Maya can continue working and focus on other activities.

To help Maya remember where the virtual objects have been located, sMoRe projects a 2.5D layout map on one of the room's walls. This map displays small low-poly versions of all the virtual objects in her apartment in the exact relative locations. This simplified representation provides an intuitive overview of her space, allowing Maya to easily understand the arrangement of items. The objects on the map are fixed and cannot be interacted with. After comprehending and exploring the space with the layout map, Maya notices that a low-poly version of the car key has also been generated on the layout map at the exact relative location of the virtual car key. 

After creating the car key's visual reminder, Maya begins sketching digital objects using sMoRe. She starts creating a 3D soccer ball, generating several prompts for rendering the ball. sMoRe listens and provides four different options. Maya can request new versions and continue exploring. Once she selects her preferred version, sMoRe asks where she would like to place it. When Maya responds, \textit{"on my desk,"} sMoRe positions the virtual ball accordingly. Since the system is aware of the room's physical boundaries and geometry, the virtual ball remains fixed on the desk, allowing Maya to move around without altering its position. This allows her to look at the virtual objects from different angles and walk in the room normally. 

Using her controller, she can grab and move the virtual objects around the room. For example, she can move the virtual car key or the ball to a different position. Maya finds out that the virtual car key will not intersect or touch any physical objects or physical objects in the space. When she tries to drag the key passing through another object, physical or virtual, the virtual object will stop moving at the point of collision. Additionally, as she moves the virtual car key in the physical space, the layout map updates in real-time to reflect the new position. She can also interact with the low-poly car key on the layout map, in which case the 3D car key moves accordingly in the physical space. After placing several virtual objects as visual reminders at desired locations, Maya can easily recall where important items are located in her apartment by referencing the virtual objects in both the physical space and the layout map. 

\begin{figure}[!htb]
    \centering
    \includegraphics[width=1\linewidth]{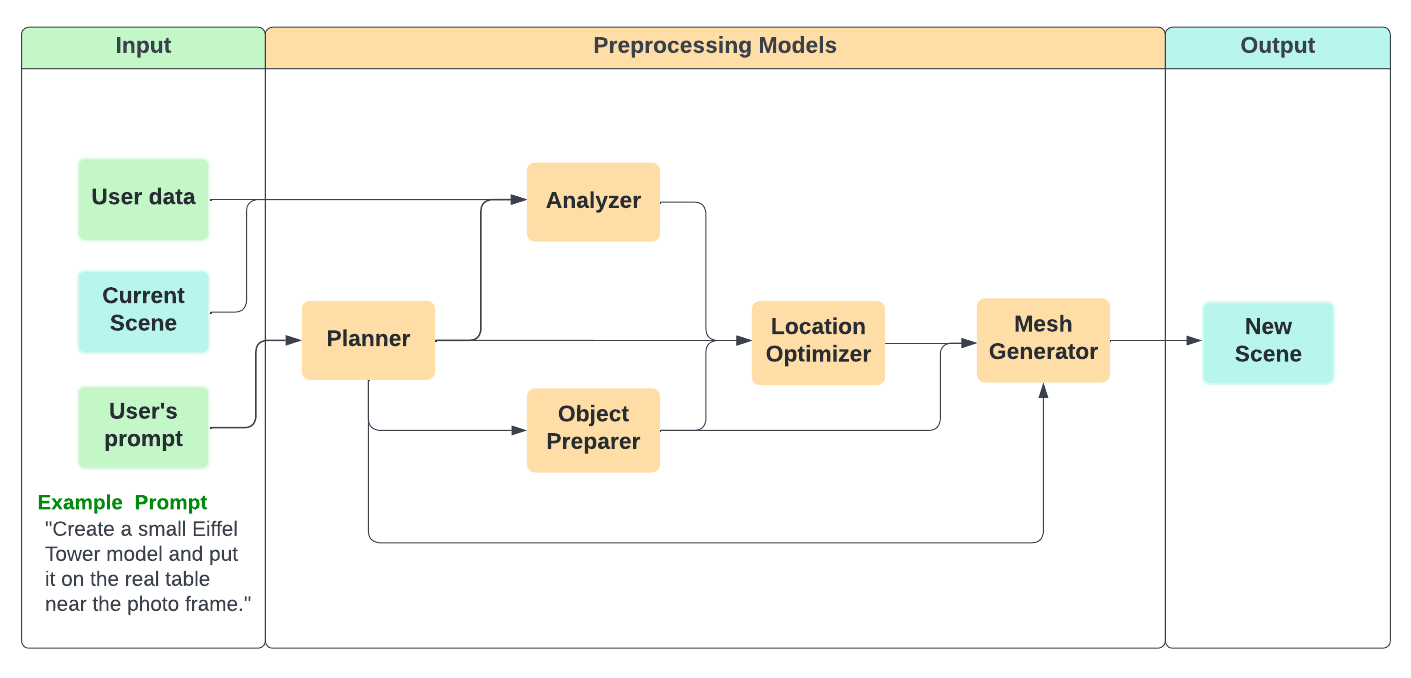}
    \caption{The workflow of sMoRe}
    \label{fig:workflow}
\end{figure}

\subsection{System Architecture}
sMoRe generates 3D virtual objects and places them at desired locations based on users' voice or text input using LLM models. Each model is guided by specialized prompts that serve different roles, such as analyzing spatial information or generating object meshes, ensuring accurate interpretation of commands and spatial actions. Users can provide commands either through voice in the MR environment or by typing prompts using an on-screen keyboard. Pseudo-code \ref{ps:workflow} shows the overall workflow of sMoRe, from detecting the user' prompt to generating the virtual object.

The system takes the user's prompt $\pi$, the user's current scene information $\Omega$, and the user's context $\upsilon$ (which includes the user's position $p_u$ and orientation $o_u$) as inputs. sMoRe uses this information and outputs a virtual object $O$ that will be placed at the user's desired location, along with other properties like dimensions and physics.

As illustrated in Fig \ref{fig:workflow}, if users opt to use voice commands, sMoRe will first transcribe the speech into text using a speech-to-text conversion service. Once the prompt is extracted, then it is passed to the \texttt{Planner}. The system uses $\pi$ to detect the target virtual object $\tau$ to generate in the MR environment, extracting the name, description, and relationships with existing locations from the users' prompt. Locations could refer to either references to objects (e.g., ``on the table'') or a specific space in the environment (e.g., ``in the middle of the room'').

The target virtual object $\tau$, the scene information $\Omega$, and the user context $\upsilon$ are provided to the \texttt{Analyzer}, which links the current scene's physical locations and physical objects with the locations and objects described and embedded in $\tau$. The \texttt{Analyzer} outputs a set of tuples $R_\tau$ in which each tuple represents a link between physical and virtual locations.

In parallel, the \texttt{Object Preparer} uses $\tau$ to define the graphic properties of the target object $P_\tau$. These properties include the shape, dimensions $(w, h, d)$, and physical characteristics, such as gravity and mass. These properties are defined based on either user specifications or real-world references suggested by an LLM.

Once $\tau$, $R_\tau$, and $P_\tau$ are ready, the \texttt{Location Optimizer} determines the optimal location for the virtual object within the scene $\Omega$ based on the detected locations and objects. It then instantiates a placeholder object $L_\tau$ with the specified properties at the optimal location.

Then, a module called \texttt{Mesh Generator} takes $\tau$ and employs a text-to-3D generative AI model to create the final mesh for the virtual object (i.e., its detailed geometric representation). The system creates the final object $O$ using $P_\tau$ to determine the final object's physical attributes and $L_\tau$ to locate the final object within the scene, completing the integration into the scene.

This workflow allows users' instructions to be effectively translated into well-defined virtual objects, placed accurately in the mixed reality environment, and presented with realistic properties and interactions. In the following subsections, we describe each component's function in detail.

\begin{algorithm}
\caption{sMoRe Workflow for Virtual Object Generation and Placement}
\begin{flushleft}
\textbf{Input:} User's prompt $\pi$, scene information $\Omega$, user context $\upsilon$ (position $p_u$, orientation $o_u$)\\
\textbf{Require:} Planner $P(\cdot)$: Task breakdown module;\\
\hspace*{3.5em} Analyzer $A(\cdot)$: Scene analysis module;\\
\hspace*{3.5em} Object Preparer $OP(\cdot)$: Object property definition module;\\
\hspace*{3.5em} Location Optimizer $LO(\cdot)$: Object placement module;\\
\hspace*{3.5em} Mesh Generator $MG(\cdot)$: Text-to-3D generative AI model.
\end{flushleft}

\begin{algorithmic}
\State $\tau \gets P(\pi)$ \Comment{Decompose the prompt $\pi$ into manageable tasks $\tau$.}
\State $R_\tau \gets A(\tau, \Omega, \upsilon)$ \Comment{Analyze the scene to identify relevant locations.}
\State $P_\tau \gets OP(\tau)$ \Comment{Define the properties of the virtual object.}
\State $L_\tau \gets LO(\tau, R_\tau, P_\tau, \Omega)$ \Comment{Determine the optimal location for the placeholder object.}
\State $O \gets MG(\tau, L_\tau, P_\tau)$ \Comment{Generate the final 3D mesh for the virtual object.}
\end{algorithmic}
\label{ps:workflow}
\end{algorithm}

\begin{figure}[!htb]
    \centering
    \includegraphics[width=0.8\linewidth]{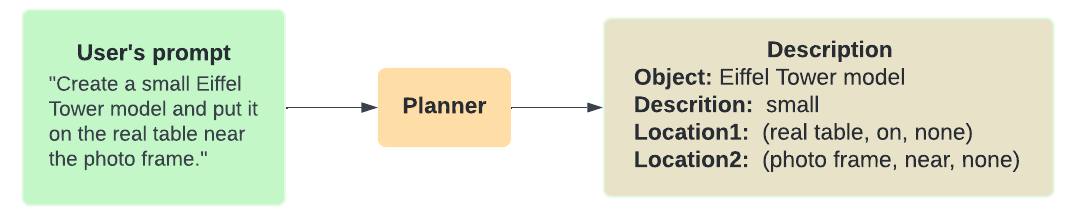}
    \caption{Example of the \texttt{Planner}}
    \label{fig:planner}
\end{figure}

\subsubsection{Planner}
As shown in Fig \ref{fig:planner}, the system employs the \texttt{Planner} module, powered by a LLM, to transform the user's prompt \(\pi\) into the target object \(\tau\) to be placed in the MR environment. To effectively generate and place virtual objects, the system must accurately interpret the user's input and extract the necessary information.

The \texttt{Planner} LLM component is prompted process \(pi\), breaking it down into essential descriptions. Specifically, it extracts key information required for generating and placing the virtual object. The LLM component processes the users' prompt \(\pi\) as follows: $P(\pi) \rightarrow \tau = (n, a, \mathcal{L})$, where \(\tau\) includes $n$, the name of the object (e.g., \textit{"car key"}), $a$, the detailed attributes of the object (e.g., \textit{"large"} or \textit{"upside down"}), and a set of locations $\mathcal{L}$. Each location $l$ in $\mathcal{L}$ can be represented as \(l= (to, sr, ad)\), where $to$ is the location information (e.g., \textit{"the table"}), spatial relation $sr$ with respect to $\tau$ (e.g., \textit{"next to"}), and any additional details of the location $ad$ (e.g., \textit{"\textbf{a little} far from the table"}).


\begin{figure}[!htb]
    \centering
    \includegraphics[width=1\linewidth]{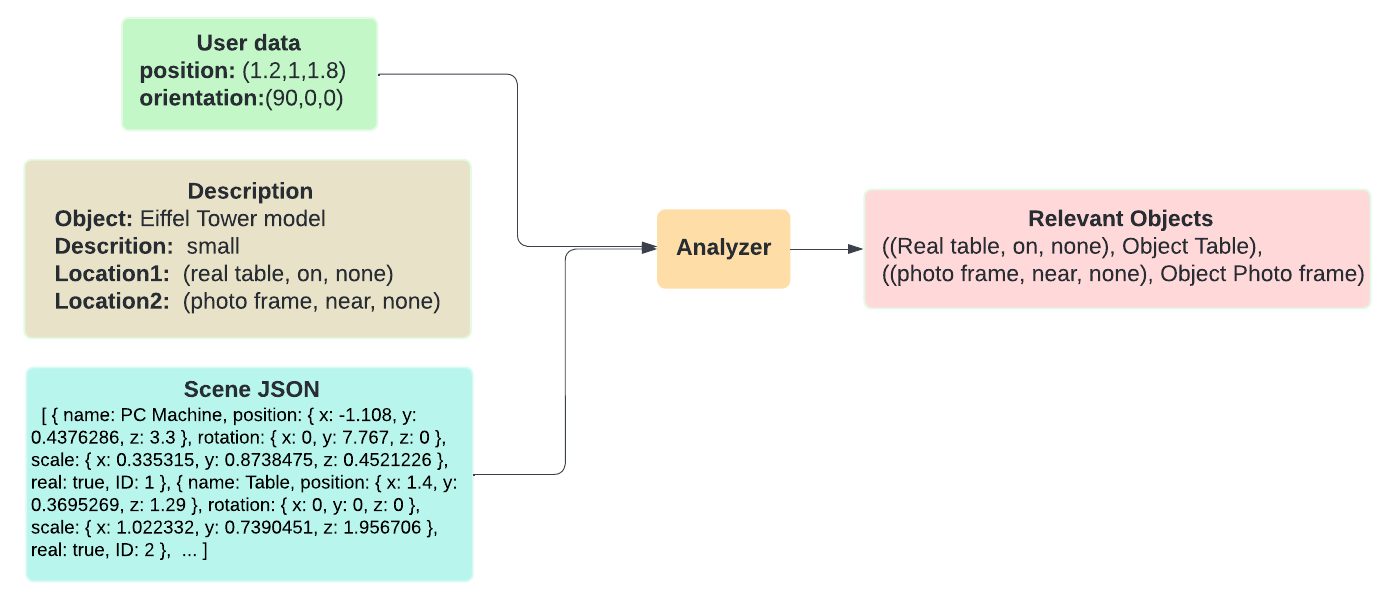}
    \caption{Example of the \texttt{Analyzer}}
    \label{fig: analyzer}
\end{figure}

\subsubsection{Analyzer}
To generate interactive 3D scenes in MR, sMoRe needs to map the relationships between the physical environment and the virtual objects. The goal of the \texttt{Analyzer} is to identify locations relevant to users' commands by processing the target object \(\tau\), the scene information \(\omega\), and the user context \(\upsilon\), which includes the user's current position \(p_u\) and orientation \(o_u\)).

The \texttt{Analyzer} takes these inputs to determine the relevant locations in the scene. Specifically, the scene $\Omega$ includes detailed information about all objects in the 3D digital environment, such as their names, positions, rotations, scales, and whether they are virtual or physical. We represent the information of $\Omega$ using a JSON file generated automatically by the system based on a 3D scan of the room in FBX format, which is a widely used format for representing complex 3D models.

If the target object's location is based on the user's position (e.g, \textit{"in front of me"}), the \texttt{Analyzer} uses the user context $\upsilon$, including the user's current position and orientation (\(p_u\), \(o_u\)), to create a user object $O_u$ as the relevant object. As shown in Fig \ref{fig: analyzer}, the \texttt{Analyzer} module, denoted as \(A(\tau, \Omega, \upsilon)\), outputs the relevant locations \(R_\tau\), represented as $R_\tau = \{(l_1, O_1), (l_2, O_2), \dots\}$, where each pair \((l_i, O_i)\) includes the target object's location \(l_i\) and the corresponding object \(O_i\) stored in $\Omega$, along with information such as position, rotation, and scale.

\subsubsection{Object Preparer}
\begin{figure}[!htb]
    \centering
    \includegraphics[width=0.8\linewidth]{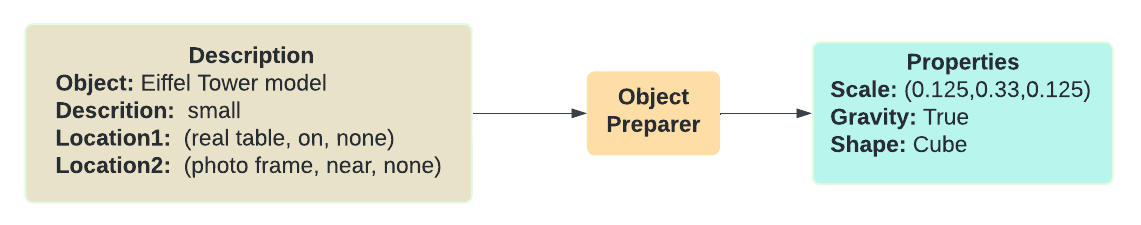}
    \caption{Example of the \texttt{Object Preparer}}
    \label{fig:preparer}
\end{figure}

As illustrated in Fig \ref{fig:preparer}, This module helps the system define the dimensions of the target object $\tau$ to align them with real-world dimensions and physics. It processes the target object \(\tau\) to define its physical properties \(P_\tau\), including shape, dimensions \((w, h, d)\), and physical characteristics (e.g., gravity and mass) based on user specifications or real-world references provided by an LLM. Here are the properties considered by this module's LLM:

\begin{itemize}
    \item \textit{Scale:} The LLM pre-defines the overall size of the object to match real-world proportions or user-defined dimensions. If the descriptions of the objects are extracted by the \texttt{Planner}, they are incorporated in the prompt to guide scaling. Otherwise, the system prompts an LLMs to set the scale to approximate real-world size, making the object proportionate and fitting naturally within the mixed reality environment. Boundaries (i.e., colliders in Unity) are generated automatically based on the scale, making correct scaling crucial for creating precise boundaries for the \texttt{Location Optimizer}.
    \item \textit{Rigidbody:} A Rigidbody component in Unity allows objects to be affected by physics, including gravity and kinetic interactions. To simulate realistic physics, the \texttt{Object Preparer} component configures the Rigidbody settings dynamically, enabling or disabling gravity depending on the object's nature (e.g., disabling gravity for a "balloon") or specific user instructions (e.g., if the user specifies "put this on the wall").
    \item \textit{Shape:} The shape property defines the geometry of the virtual object, such as whether it is a cube, sphere, cylinder, or a custom shape. Defining the shape is helpful for the \texttt{Location Optimizer} to accurately detect the optimal location for placement. The \texttt{Object Preparer} selects the shape based on the user's description or known real-world references by their names. 
\end{itemize}

The module's LLM component outputs these properties for the object, represented as: \(P_\tau = \{w, h, d, \text{physics}, \text{shape}\}\). The \texttt{Object Preparer} then instantiates an object with the detected geometrical shape in Unity with the specified scale, rotation, and Rigidbody settings. This cube is subsequently passed to the \texttt{Location Optimizer}, which uses it to determine the optimal placement that meets the user's requirements.

\subsubsection{Location Optimizer}
\begin{figure}[!htb]
    \centering
    \includegraphics[width=0.8\linewidth]{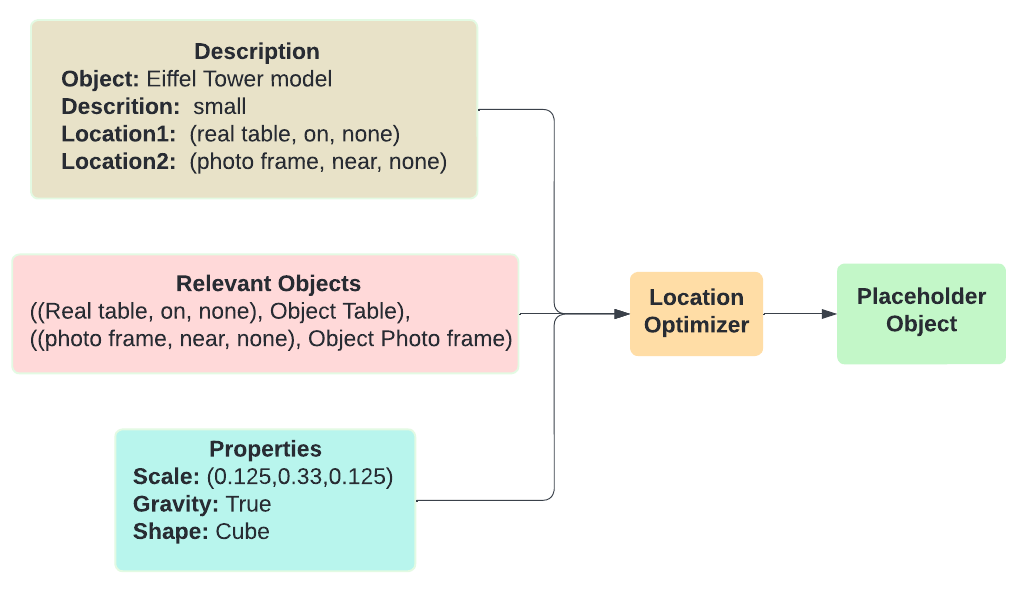}
    \caption{Example of the \texttt{Location Optimizer}}
    \label{fig: location optimizer}
\end{figure}

As shown in Fig \ref{fig: location optimizer}, This module is responsible for analyzing and determining the best placement for the target object $\tau$ (\textbf{DG2}). To ensure that $\tau$ is positioned according to users' voice commands while satisfying spatial constraints, the system employs a structured approach leveraging the Mixed Reality Utility Kit (i.e., MRUK). Pseudo-code \ref{ps:location} provides more details of how the locations are determined.

Given the target object $\tau$, its set of relevant locations (\(R_\tau\)), and its physical properties (\(P_\tau\)), the \texttt{Location Optimizer} renders the placeholder object (\(L_\tau\)) in the scene $\Omega$ at the optimal position $(x,y,z)$. To start, this model uses the MRUK's pre-defined script to generate random points on specific surfaces within the scene. The script enables the system to define the surface type and generate random points accordingly. Surface types include vertical, on top, floating, hanging down, or any combination thereof. The module's LLM component identifies the appropriate surface type for the target object (\(\tau\)) provided by the user. For example, if the user specifies ``on the wall,'' the LLM interprets the command and returns the vertical surface type, and the system adjusts the point generation to match. This step is important to ensure objects are placed as intended, whether on a wall, table, or ceiling.

\begin{algorithm}
\caption{Location Optimizer for Object Placement}
\begin{algorithmic}
\State \textbf{Input:} Target object $\tau$, set of relevant locations $R_\tau$, and physical properties $P_\tau$
\State \textbf{Output:} placeholder object $L_\tau$

\For{each attempt until Maximum attempt}
    \State Determine surface type based on the target object $\tau$ using LLM interpretation
    \State Generate random candidate point $p_c = (x, y, z)$ on the identified surface
    \State Transform $p_c$ into candidate location object $L_c$ with properties $P_\tau$
    \For{each location $l_i$ and corresponding object $O_i$ in $R_\tau$ }
        \State Validate candidate point according to the spatial relationship $(l_i, O_i)$
    \EndFor
    \If{$L_c$ is valid}
        \State $L_\tau = L_c$
        \State \textbf{break}
    \EndIf
\EndFor

\If{no valid point found after maximum attempts}
    \State Default to placing $L_\tau$ on the floor in front of the user
\EndIf
\end{algorithmic}
\label{ps:location}
\end{algorithm}

Once a candidate point $p_c = (x, y, z)$ is generated, it is transformed into a candidate location object $L_c$ that includes properties derived from $P_\tau$ and $p_c$. This transformation allows the system to consider both the physical attributes and the spatial constraints relevant to the target object.

The system immediately validates each candidate location object $L_c$ before deciding whether to generate another one. The validation process checks whether the candidate object meets the conditions for each spatial relationship $l_i$ with respect to the target object $O_i$. Specifically, it verifies if the location object $L_c$ fulfills the requirements for the spatial relationship $l_i$, such as positioning on the correct surface or maintaining the appropriate distance from the target object.

To handle different spatial relationships, the system combines pre-defined rules with LLM interpretation. Pre-defined rules ensure accurate placement for typical relationships. For instance, if the relation is "on," the system checks whether the candidate location object $L_c$ falls within the boundary of the target object along both the $x$ and $z$ axes. The LLM is used to match similar terms and interpret nuanced relationships. For example, terms like "next to" or "near" are recognized as equivalent, and the LLM applies a consistent validation rule. Table \ref{table:simple-spatial-relations} presents all the pre-defined rules for valid spatial relationships. Due to limitations in handling colliders and meshes in Unity, the "inside" relationship is not currently supported.

The system continues this process until a valid candidate LOCATION object $L_c$ is found or until the pre-set maximum number of attempts is reached. If a valid point is found, $L_\tau$ will be set to be $L_c$. If no valid point is found after all iterations, the system defaults to generating a point on the floor in front of the user. This fallback approach ensures that the object is still placed in the scene, even when ideal conditions are not met.

\begin{table}[!htb]
\centering
\caption{Spatial Relationship Criteria}
\begin{tabular}{|c|c|}
\hline
\textbf{Spatial Relation} & \textbf{Criteria} \\ \hline
On       & Within $x$ and $z$ boundaries \\ \hline
Left Of   & Completely to the left side (negative $z$ direction) \\ \hline
Right Of  & Completely to the right side (positive $z$ direction) \\ \hline
Next To   & Close in both $x$ and $z$ directions without overlapping \\ \hline
In Front Of & Directly in front (negative $x$ direction) \\ \hline
Behind   & Directly behind (positive $x$ direction) \\ \hline
Above    & Higher up in the $y$ direction \\ \hline
Below    & Lower down in the $y$ direction, aligned \\ \hline
Far From  & Far away in both $x$ and $z$ directions \\ \hline
Outside  & Completely outside the boundaries of another object \\ \hline
Aligned With & Aligned along the same $x$, $y$, or $z$ axis \\ \hline
Touch  & Surfaces of objects are in contact \\ \hline
\end{tabular}
\label{table:simple-spatial-relations}
\end{table}
\begin{figure}[!htb]
    \centering
    \includegraphics[width=0.8\linewidth]{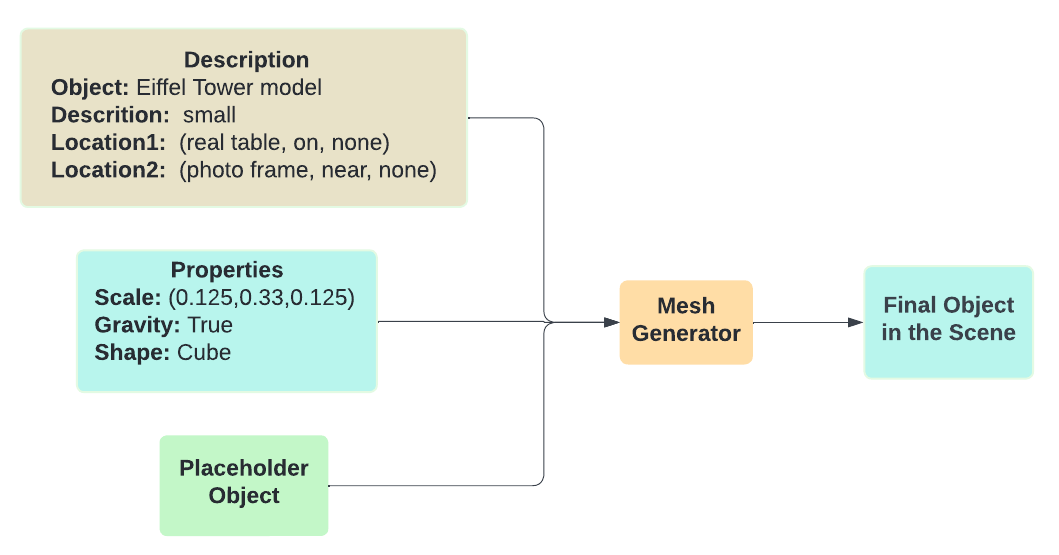}
    \caption{Example of the \texttt{Mesh Generator}}
    \label{fig: mesh generator}
\end{figure}

\subsubsection{Mesh Generator}
Given the target object (\(\tau\)), the location object (\(L_\tau\)), and the physical properties $P_\tau$, the \texttt{Mesh Generator} renders the final virtual object (\(O\)) for the scene $\Omega$. This module serves as the creator module and is responsible for generating new virtual objects. 

We integrated a text-to-3D generative AI module named \textit{Genie}\footnote{\url{https://lumalabs.ai/genie}} for creating these objects. By sending the text description of the desired object (i.e., the name \(n\) and detailed description \(a\)) to \textit{Genie}, the \texttt{Mesh Generator} receives four 3D object options. These four options are displayed as short-loop video previews on a 2D screen in front of the user's view. Once the user clicks on one of the video previews, the selected object is downloaded and passed to the \texttt{Mesh Generator}.

Afterward, the \texttt{Mesh Generator} attaches additional components to the generated object to ensure it functions properly in the mixed reality environment:
\begin{itemize}
    \item \textit{Boundary:} A collider is automatically assigned to each object, allowing for proper interaction with the environment. Specifically, a mesh collider is used to accurately match the object's shape, ensuring realistic physics interactions and collision detection.
    \item \textit{Interaction Properties:} Interactive components are attached to each object, enabling users to pick up, move, and manipulate the objects using their controllers in the mixed reality environment. Further details regarding these interaction properties are discussed in Section \ref{sec: Interactive Features}.
\end{itemize}

\begin{figure}[!htb]
    \centering
    \includegraphics[width=1\linewidth]{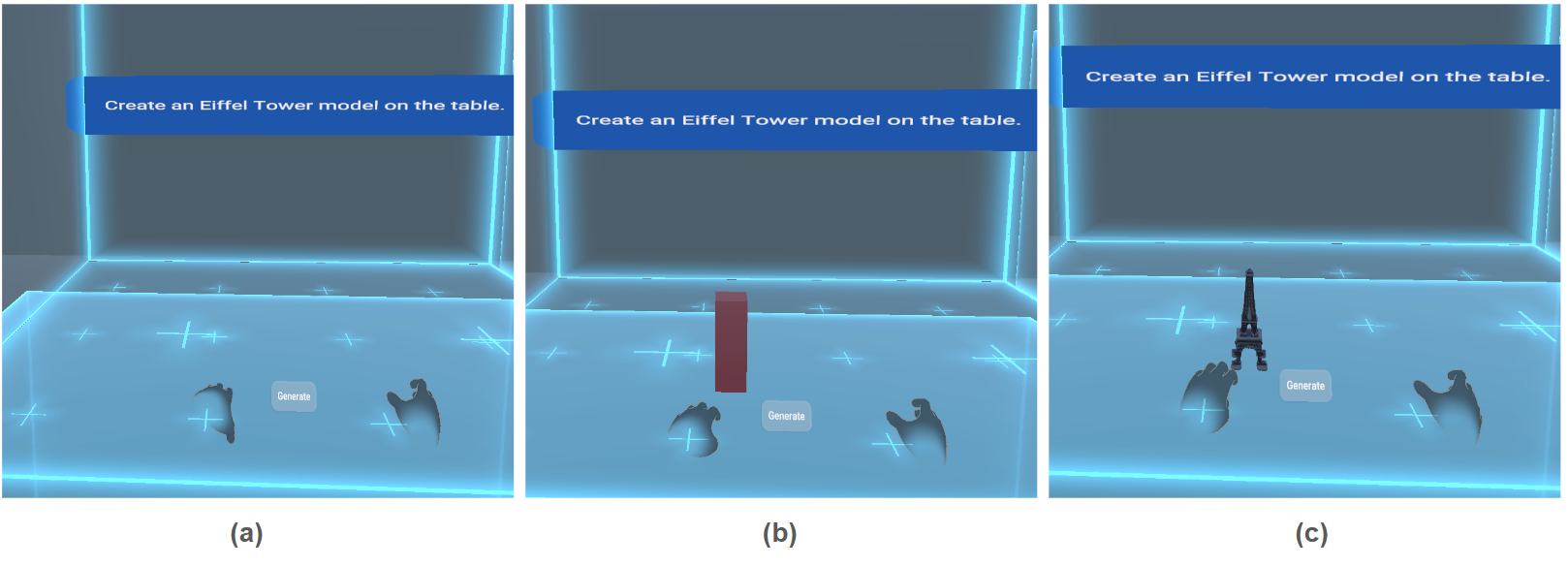}
    \caption{The procedure of generating the target object (\(O\)). Since Meta Quest Oculus 3 does not provide access to the user's point-of-view (POV), we display the blue boundary of a table for demonstration purposes. (a) The user creates the prompt and requests sMoRe to generate the object. (b) A placeholder object ($L_\tau$) shows up on the table. (c) The target object is generated and replaces the placeholder object.}
    \label{fig:target-object}
\end{figure}

Finally, the \texttt{Mesh Generator} matches the scale, rotation, and position of the generated object (\(O\)) to that of the placeholder object (\(L_\tau\)). Once the final object is correctly scaled and positioned, the placeholder (\(L_\tau\)) is deleted, leaving the fully realized virtual object in the scene. Fig \ref{fig:target-object} shows how sMoRe generates the placeholder (\(L_\tau\)) and the final target object (\(O\)).

\subsection{Additional features}
\label{sec:key-features}
\subsubsection{2.5D Layout Map}
\begin{figure}[!htb]
    \centering
    \includegraphics[width=0.6\linewidth]{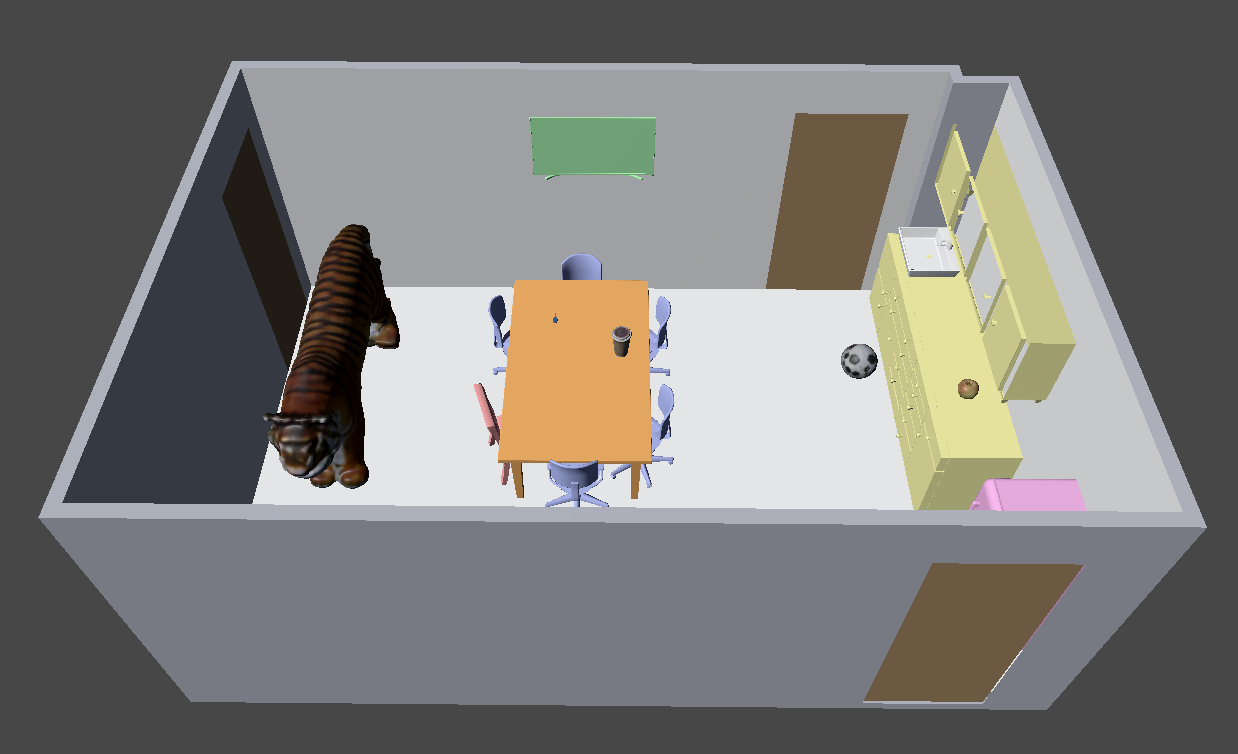}
    \caption{An example layout map of a conference room with some randomly generated objects.}
    \label{fig:layout}
\end{figure}

To assist users in better understanding the layout of the space and objects within it (\textbf{DG1}), sMoRe provides a vertical 2.5D Layout Map displaying a bird's-eye view of the room. This view combines both a top-down (2D) layout and a slight 3D angle, allowing users to perceive the spatial relationships and depths of various objects in the environment. Fig \ref{fig:layout} illustrates an example of the 2.5D layout map. Upon entering the room, the layout is projected on a wall in a fixed size, showing the layout of the scanned physical objects in the room. When creating a virtual object in the MR environment, a low-poly and smaller version of the object will be generated on the layout map at the location corresponding to the relative position of the original virtual object in the room. 

The low-poly layout objects show the physical shape of the original virtual objects and have name labels attached to their bodies to make them easily recognizable. Additionally, using simplified geometry in the low-poly models reduces visual clutter, making the overall layout easier to comprehend at a glance. Other than the location, the scale of the layout objects also corresponds to the scale of the original virtual objects, which helps users to recognize the relative sizes of the objects quickly. Furthermore, the physical relations between virtual objects in the room are maintained on the map. For example, if there is a plate on the table, the 2.5D layout view would show the plate slightly elevated above the table's surface to represent its actual position. In contrast, if the plate is under the table, besides indicating the relative position, the system would use a semi-transparent outline or a dotted line to represent the plate's boundary to hint at its position beneath the table. The semi-transparent outline would help users understand that the plate is under the table without fully revealing it, preserving the spatial relationship between the two objects.

\subsubsection{Hand/Controller Interaction}
\label{sec: Interactive Features}
To further support \textbf{DG3}, sMoRe offers controllers and hand tracking as the primary interaction methods. By performing finger pinch gestures or pressing the controller's grip button, users can grab virtual objects and move them to the desired location. The system sets a mesh collider to each object, so when grabbing and moving objects, they do not overlap with the existing objects. When the grabbed objects collide with a virtual object, the virtual object would react based on physics, such as bouncing, sliding, or toppling over. When the grabbed objects collide with a physical object, the object will stay stationary, as it is part of the real world and not affected by the virtual forces applied in the mixed reality environment. sMoRe also supports distance grabbing, so users can interact with virtual objects from a distance by pointing at them and performing grabbing using controllers or hands.

Besides the virtual objects, users can also interact with layout objects on the 2.5D layout map. These layout objects can only stay within the layout map fixed on the wall. By grabbing and moving the layout objects on the map, the corresponding virtual objects would also move based on the relative position on the map. Similarly, the layout objects corresponding to physical objects stay stationary. To synchronize the layout objects and virtual objects in the scene, we follow the formula:

\[
\mathbf{P}_{\text{virtual}} = \mathbf{P}_{\text{origin}}^{\text{virtual}} + S_{\text{virtual}} \cdot \left( \mathbf{P}_{\text{layout}} - \mathbf{P}_{\text{origin}}^{\text{layout}} \right)
\]
\[
\mathbf{P}_{\text{layout}} = \mathbf{P}_{\text{origin}}^{\text{layout}} + \frac{1}{S_{\text{virtual}}} \cdot \left( \mathbf{P}_{\text{virtual}} - \mathbf{P}_{\text{origin}}^{\text{virtual}} \right)
\]

Where \( \mathbf{P}_{\text{virtual}} \) is the position of the virtual object in the 3D space. \( \mathbf{P}_{\text{layout}} \) is the position of the layout object on the 2.5D map. \( \mathbf{P}_{\text{origin}}^{\text{virtual}} \) is the origin point of the virtual space in 3D. \( \mathbf{P}_{\text{origin}}^{\text{layout}} \) is the origin point of the layout map on the 2.5D map. \( S_{\text{virtual}} \) is the scaling factor between the virtual space and the layout map.


\subsection{Implementation details}
We developed sMoRe in Unity 2022.3.20f1 and C\#. For user voice transcription, we utilized Microsoft Azure's standard-tier speech-to-text service. In detail, we employed the SpeechRecognizer class provided by the Microsoft Cognitive Services Unity API. By fine-tuning the SpeechRecognizer's recognized event, we processed every sentence from users' speech in real-time. We set the segmentation silence timeouts to 0.3s. We set all the other parameters to default. 

After the system detects a sentence, the recognized text is displayed on a blue screen in front of the user within the mixed reality environment. This visual feedback allows the user to confirm the detected text before proceeding. Additionally, a button appears on the blue screen, enabling the user to initiate object generation based on the detected sentence by either using a controller or performing a poking gesture.

If the user prefers to type their instruction instead of using voice commands, they can do so by performing a left-hand pinch-palm gesture to summon a virtual keyboard. Users can type their desired command using the virtual keyboard and press the enter button to generate the object. The keyboard will disappear if the pinch-palm gesture is performed again.

We integrated Microsoft Azure OpenAI's GPT-4o model for sMoRe. For environment scanning, we utilized the LiDAR scanning software \textit{PolyCam}, which allows users to scan their physical room with objects. The scanned room data is imported into Unity in FBX format, ensuring that the virtual environment accurately matches the physical space. We set a maximum iteration value of 10,000 for the \texttt{Location Optimizer} to allow a valid placement to be found without causing significant delays.

\section{User Study}
\label{user_study}
We conducted a laboratory user study with six participants in October 2024 to evaluate this system. We asked participants to assess sMoRe's usability, effectiveness, and usefulness. The study was reviewed and approved by our institution's IRB office (\#23-09-8107). We aim to answer the following questions:

\begin{itemize}
    \item \textbf{RQ1:} Can users effectively and easily utilize sMoRe to place and manipulate virtual objects within the mixed reality environment as intended? 
    \item \textbf{RQ2:} How useful is sMoRe in enhancing users' ability to understand and interact with both virtual and physical elements within their space?
    \item \textbf{RQ3:} What advantages and challenges do users perceive when using sMoRe to generate virtual objects?
\end{itemize}

\subsection{Recruitment and Participants}
We recruited six participants from a private university in the US. The participants were all graduate students. All participants were fluent in English. In terms of gender, five were male students, and one was a female student. Participants volunteered for this study and were compensated with a \$15 gift card for their time. 

\subsection{Procedure}
Each study session took approximately 30 minutes and was conducted in person in our research laboratory. At the beginning of each session, the research assistants explained the purpose of the study and collected informed consent forms from the participants. Participants then completed a pre-treatment questionnaire that assessed their demographic information and familiarity with MR. Once the participants finished the survey, the research assistant guided each participant to a room with a Meta Oculus Quest 3 device connected to a computer that rendered the sMoRe application. The participants adjusted the headsets and started using the system with a brief tutorial explaining how to control the environment. 

In the MR environment, sMoRe showed the participants its key features and explained the interaction methods that they could use. Each participant took two minutes to explore the system using the instructions shown in the application. After familiarizing themselves with the interactions, they were asked to give voice commands to generate objects and place them at their desired locations. Each participant took ten minutes to explore generating virtual objects and interact with the objects using their controllers and hands. 

After the session was completed, participants responded to a final questionnaire to report their experiences and opinions about the system. This final questionnaire included NASA's Task Load Index (TLX), which assesses workload on a 7-point scale \cite{hart1988development}, and the SUS usability \cite{brooke1996sus} scale that assesses a system's usability characteristics based on ten items using a 5-point scale. We also designed 5-point Likert scales to measure participants' satisfaction with the app, its effectiveness, its usefulness, and whether they will use it again. We also asked participants to respond to open-ended questions that addressed the advantages, obstacles, and suggestions for sMoRe.

\begin{figure}[!htb]
    \centering
    \includegraphics[width=0.9\linewidth]{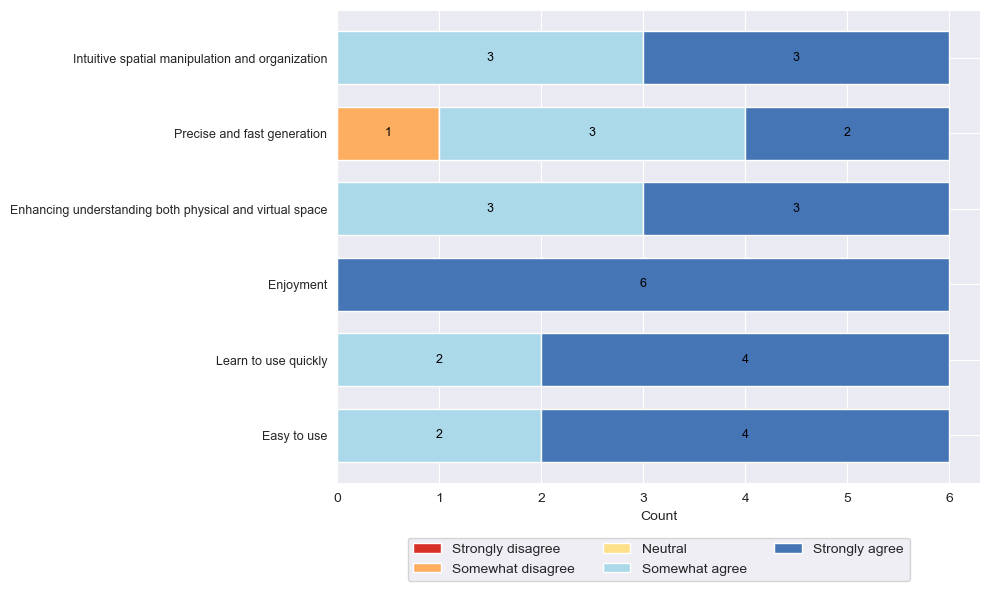}
    \caption{Selected Results from the post-study questionnaire}
    \label{fig:results}
\end{figure}

\subsection{Results}
\begin{figure}[!htb]
    \centering
    \includegraphics[width=0.8\linewidth]{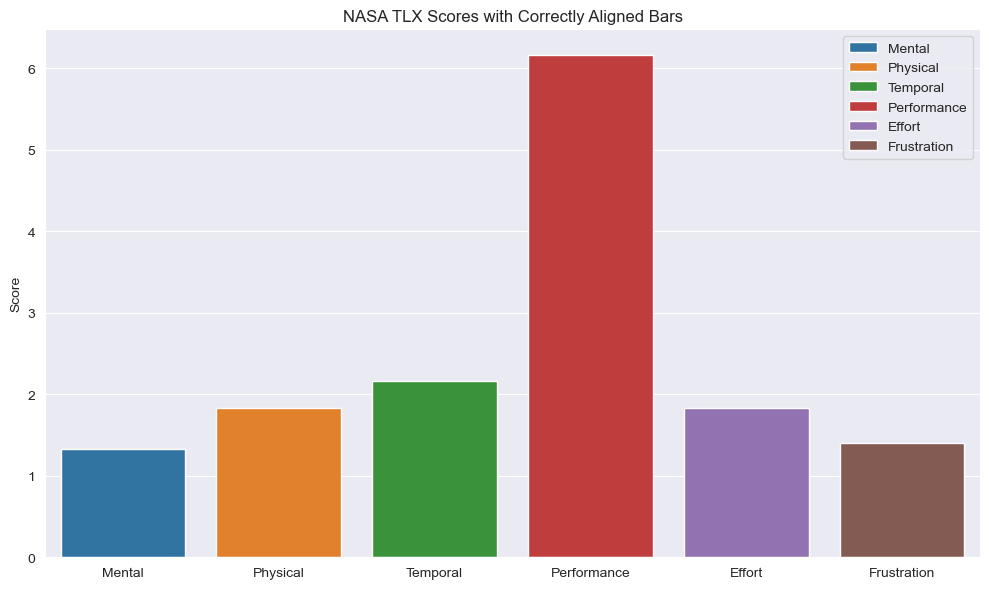}
    \caption{NASA TLX Results from the post-study questionnaire}
    \label{fig:nasa-tlx}
\end{figure}

Participants' responses are presented in Fig \ref{fig:results} and Fig \ref{fig:nasa-tlx}. According to the NASA TLX items, the majority of the participants agreed that using sMoRe for this task was not physically demanding ($M=2.33, SD=0.75$) and not highly mentally demanding ($M=2.33, SD=0.75$). Moreover, most of the participants reported feeling successful in accomplishing the task ($M=5.67, SD=1.25$). Regarding sMoRe's usability, most participants found the system easy to use ($M=4.5, SD=0.5$) and imagined most people would learn how to use the system quickly ($M=4.83, SD=0.37$). They did not find too much inconsistency ($M=2.00, SD=1.0$) or unnecessary complexity ($M=2.17, SD=1.34$) in the system. Moreover, participants found that they enjoyed the sMoRe system a lot ($M=4.83, SD=0.37$). The majority of the participants were satisfied with the virtual objects created by the system ($M=4.5, SD=0.5$) and using the MR environment ($M=4.33, SD=0.47$). Finally, participants were able to achieve what they had in mind with sMoRe ($M=4.5, SD=0.5$). 

Participants found creating voice prompts to generate virtual objects in the scene simple and intuitive. P4 stated, ``I speak directly to the system like talking to a human being.'' P3 wrote that ``I used a simple sentence to indicate the object I would like to create and the location of the object that I would like to place'' and ``The location would be described as a place relative to the furniture already in the room, which is also in the AR/MR world.''

Comparing their typical experience of manipulating a 3D environment, participants described sMoRe as ``different,'' ``intriguing,'' and ``easy to use.'' P1 noted that the ability to enter commands via speech felt ``new and different'' compared to previous 3D creation experiences. P3 stated, ``My prior experience with 3D object generation was slow and needed a lot of construction from scratch and fine-tuning,'' and needs ``a lot of prior knowledge and skills of the 3D creation software and art skills.'' Thus, P3 appreciated that sMoRe only required ``simple sentences'' to create objects and specify ``locations relative to the furniture already in the room.'' P5 found the system to be ``straightforward'' and praised its capability to generate ``novel and unimaginable'' objects, adding that this feature made the interaction ``more creative'' and engaging. P4 mentioned that the system was ``pretty new to me'' but also expressed a desire to use it for ``fun'' during their spare time. P6 highlighted that the system's approach to generating and interacting with virtual objects was ``easy to use'' and ``interesting,'' despite mentioning a slight lag in hand gesture recognition compared to other systems like Apple Vision Pro. Overall, participants emphasized that sMoRe allowed them to interact naturally and effectively with virtual environments.

Despite these advantages reported by the participants, others noted several challenges with the system. P1 mentioned that ``the placement of the object was hard,'' particularly when generated items lacked a firm structure, making them difficult to position or stabilize. P2 highlighted that ``the object generation process took longer than expected'' and that it was challenging to precisely locate objects for interaction. P3 reported that ``the location description by natural language was not accurate enough'' and difficulties with hand gesture manipulation. P4 expressed that the conversion of instructions to text was ``too sensitive,'' leading to difficulties when prompts were not recognized accurately, and noted that ``the interaction part still has space to improve,'' especially regarding hand gesture recognition. P5 mentioned that when attempting to grab an item, ``the item will be stuck to my finger, and it is hard to throw away,'' suggesting a need for more refined handling mechanics. P5 also suggested a ``better database for the generated objects'' to improve object variety and quality. These challenges indicate areas where sMoRe could be enhanced to improve user interaction and overall experience.

\section{Discussion}
\label{discussion}
In this study, we present sMoRe, a novel MR application designed to create virtual objects using GenAI and locate them where users want by using LLMs. We addressed the challenge of object creation and placement using advanced AI models, providing users with a more interactive and fluent experience with MR. In this section, we delve deeper into the implications of this study, analyzing how sMoRe can contribute to enhancing MR applications.

First, we designed sMoRe to extend the current MR capabilities for generating and placing virtual objects. This approach can potentially reduce users' cognitive and physical demands associated with manipulating 3D objects in XR environments. While previous systems rely on controllers and asset libraries, sMoRe provides an advanced solution based on transforming users' prompts into realistic 3D figures using GenAI. This implementation not only simplifies the interaction process but also lowers the barrier for users who may not have prior experience with 3D modeling or MR systems. The participants' responses reinforced this point, as many of them reported having a good experience creating 3D objects and placing them in their surroundings.

Another contribution to the literature is providing a dynamic way to generate and place custom 3D objects using voice prompts in real time. One challenging aspect of this project was mapping the virtual objects, physical objects, their physical relationships, and the desired location for the object using users' voice commands. By using LLMs and scanning the user's physical room, the proposed architecture enables the system to identify which objects are already in the scene, which ones should be generated, and their physical relationships within the MR environment. This architecture augments the mere use of traditional input controllers, and enables users to use their voice to interact in a more rich way with their surroundings in MR. Despite the positive results, some participants reported that the system still made some errors when locating objects in the room. Some potential approaches to refine the system's spatial reasoning capabilities include more sophisticated environmental mapping, providing the LLM with multimodal information in addition to the user's voice command, and eye-tracking techniques.

Lastly, this study provides potential future directions for future MR systems integrated with LLM and Generative AI. For example, a collaborative version of this application could benefit teams working on creativity or problem-solving tasks \cite{Bai2020}. As a result, multiple users could build products, prototype their ideas, and reduce the use of physical resources to create tri-dimensional objects at a scale \cite{Han2024}. Moreover, multimodal interactions (e.g., gestures, drawings, gaze tracking) could provide more advanced commands to help users create virtual objects in MR environments considering their bodies' movements. GenAI could capture these cues and provide users with more ideas and virtual objects in a simple way. 

\subsection{Potential Applications}
\paragraph{Visual Reminders} One of the potential applications of sMoRe is to create visual reminders within users' physical spaces. Users can place virtual objects in specific locations to serve as reminders for tasks or items. For example, a user may place a large virtual "car key" on their cabinet to remind them where to find their actual car key, which is smaller and might be hidden within the drawer. This type of visual cue leverages mixed reality to enhance memory and organization, allowing users to effectively utilize their physical environment as a contextual reminder board. The ability to create reminders in exact locations makes the system a powerful tool for managing daily tasks without overwhelming traditional calendars or to-do lists.

\paragraph{Interior Design and Space Planning} sMoRe can also be applied to interior design and space planning. Users can visualize how new furniture or decor items would fit into their existing physical spaces before making a purchase. By generating and positioning virtual objects such as furniture, users can experiment with different layouts and decide on optimal arrangements. This approach not only saves time but also allows users to experience different spatial configurations directly in their mixed reality environment, enhancing confidence in their design decisions.

\paragraph{Creative Prototyping} For artists and designers, sMoRe offers an intuitive platform for creative prototyping. Users can create 3D sketches and virtual mock-ups of their ideas directly in mixed reality. Instead of relying on traditional 3D modeling software, which often requires technical knowledge and skills, users can simply describe their concepts using natural language and sMoRe will generate 3D models in real time. This accessibility encourages rapid prototyping and iterative design, making the creative process more fluid and user-friendly.

\subsection{Limitations and Future Work}
We acknowledge the following limitations of this work. First, the proposed system's accuracy relied on the selected LLMs when interpreting user commands. Although the system effectively handled simple and direct instructions, more complex or ambiguous inputs could lead to misinterpretations, potentially affecting object placement or generation. Future work should focus on refining the NLP capabilities to better understand nuanced user commands and handle more intricate spatial requirements.

Second, we did not test the system's scalability in handling larger and more complex MR environments. While our current implementation was tested in a small space, performance issues could arise when managing a significant number of physical objects or large environments. The model also generated one object at a time, and it was unable to generate more than one object using a single prompt. Optimization techniques, including more efficient scene processing and memory management, should be explored to improve scalability and real-time interaction in more demanding settings.

Third, the dynamic generation of 3D objects based on textual descriptions is constrained by the current capabilities of text-to-3D generative AI. We employed Genie since it was one of the most promising systems for this purpose. However, the 3D generator system can produce objects that do not fully align with user expectations, especially for highly specific or complex object descriptions. Future work should enhance the precision of object generation, potentially by integrating multi-modal AI systems that incorporate visual references or user feedback to fine-tune object design.

Conducting an extensive numerical evaluation of the system could provide more insights into the accuracy and time response of this system. Future studies should evaluate how accurate the respective matching between the predicted and the actual locations is. Moreover, testing different GenAI models and LLMs would be helpful to distinguish which models are better suited for this application.  

Finally, our user study, while demonstrating the system's effectiveness, was limited in sample size and participants' backgrounds. Future studies should involve a larger and more varied participant pool to better understand the generalizability of sMoRe's usability and the different ways it can enhance user interactions across different domains and use cases.
\section{Conclusion}
\label{conclusion}
This study introduces sMoRe, a novel MR application that combines Generative AI and LLMs to facilitate the intuitive creation and manipulation of virtual objects in physical spaces. By employing text-to-3D generation and LLMs to interpret commands and analyze spatial contexts, sMoRe provides users with a powerful, flexible tool to design and organize virtual objects in MR environments. Our user study demonstrates that sMoRe significantly enhances user interaction and interaction within MR environments, offering an enjoyable and user-friendly experience. The system's ability to automatically generate and place virtual objects based on the user's input and the physical space's constraints highlights the potential for AI-driven systems to help users engage with digital content in MR. As the interplay between AI and MR continues to evolve, applications like sMoRe represent an important step towards creating more accessible and dynamic mixed reality environments.
\begin{acks}
\end{acks}

\bibliographystyle{ACM-Reference-Format}
\bibliography{sample-manuscript}

\appendix
\section{METAPROMPTS}
In this section, we provide meta prompts for each module in sMoRe.
\subsection{Planner}
\begin{tcolorbox}[colback=blue!5!white, colframe=blue!75!black, title=Planner, breakable]
You're a helpful planner who converses with the user to come up with a plan for generating a specified virtual object in Unity. The plan will then be forwarded to another system, which will execute each step of your plan. \\ 

\# Guidelines to follow \\
- Extract the name of the virtual object users want to create from the command as well as the description of that object, focusing on physical properties like scale, rotation, etc. \\
- Attributes like color, material, or texture will be generated by AI and categorized directly under 'Objects', but if not specified, they should be omitted. \\
- When the location is not completely given, try to approximate it and follow the output format. \\
- Try to extract further details about the location relationship (e.g., 'a little', 'far from'). \\
- Print 'none' if you cannot find any. \\

\# Output format \\
Object: <name of object> \\
Description: <description of object> \\
Location 1: <location name> \\
Relation to Location 1: <relation> \\
Detail 1: <detail about relation> \\
Location 2: <location name> \\
Relation to Location 2: <relation> \\
Detail 2: <detail about relation> \\

\# Examples \\
\#\# Example \\ 
User: Please put a big white paper poster on the wall a little near me and much closer to the oven. \\

Plan: \\
Object: poster (white, paper) \\
Description: big \\
Location 1: wall \\
Relation to Location 1: on the surface \\
Detail 1: none \\
Location 2: user \\
Relation to Location 2: near \\
Detail 2: a little \\
Location 3: oven \\
Relation to Location 3: close to \\
Detail 3: much closer \\ \\

\#\# Example \\
User: Place a small glass coffee cup on the table right in front of the monitor.

Plan: \\
Object: coffee cup (glass) \\
Description: small \\
Location 1: table \\
Relation to Location 1: on the surface \\
Detail 1: none \\
Location 2: monitor \\
Relation to Location 2: in front of \\
Detail 2: right in front \\

\#\# Example
User: Put the lamp beside the couch but much closer to the window.

Plan: \\
Object: lamp \\
Description: none \\
Location 1: couch \\
Relation to Location 1: beside \\
Detail 1: none \\
Location 2: window \\
Relation to Location 2: close to \\
Detail 2: much \\

\#\# Example
User Input: Place a large green book on the shelf next to the plant and just above the lamp. \\

Plan: \\
Object: book (green) \\
Description: large \\
Location 1: shelf \\
Relation to Location 1: on the surface \\
Detail 1: none \\
Location 2: plant \\
Relation to Location 2: next to \\
Detail 2: none \\
Location 3: lamp \\
Relation to Location 3: above \\
Detail 3: just \\

\end{tcolorbox}
\subsection{Analyzer}
\begin{tcolorbox}[colback=blue!5!white, colframe=blue!75!black, title=Analyzer, breakable]
You will be given a JSON file with all white spaces and quotations removed, which contains all the game objects in a 3D Unity scene. This file will include data about each object's name, position, rotation, scale, whether it is real, and its unique ID. Additionally, you will be given the user's current position and orientation. The objects could either be real-world objects or virtual 3D objects. Your task is to interpret this JSON file, consider the user’s position and orientation, and give a description of the scene relevant to the user's request. \\

\# Guidelines to follow \\
- Relevance to Target Objects: Only summarize the part of the scene relevant to the user's request. If the user specifies a target object, provide detailed information about those objects only.\\
- Include Object Details: For each relevant object, include the name, position, rotation, scale, and other attributes (e.g., real).\\
- User's Position and Orientation: If the user is included as a target object, use their provided position, orientation, a scale of {x: 1, y: 1, z: 1}, real: true, and ID: 99. Exclude the user from the JSON file and treat them as a distinct object with these attributes. \\
- Try to extract further details about the location relationship Handling Missing Objects: If no relevant objects are found, return a message indicating this. \\
- Grouping Similar Objects: If multiple objects of the same type satisfy the user's request, group them together. \\

\# Output format \\
- Relevant Object(s): [ { name: object name, position: {x: , y: , z: }, rotation: {x: , y: , z: }, scale: {x: , y: , z: }, real: true or false, ID: number} ] \\
- If no relevant object is found: Output Relevant Object: []. \\

\# Examples \\
\#\# Example \\ 
User Input: \\
Position: (0, 0, 0) \\
Orientation: (0, 0, 0) \\
Target Objects: cup \\
Scene JSON file:
[
  { "name": "PC Machine", "position": { "x": -1.108, "y": 0.4376286, "z": 3.3 }, "rotation": { "x": 0, "y": 7.767, "z": 0 }, "scale": { "x": 0.335315, "y": 0.8738475, "z": 0.4521226 }, "real": true, "ID": 1 },
  { "name": "Table", "position": { "x": 1.4, "y": 0.3695269, "z": 1.29 }, "rotation": { "x": 0, "y": 0, "z": 0 }, "scale": { "x": 1.022332, "y": 0.7390451, "z": 1.956706 }, "real": true, "ID": 2 },
  { "name": "Cup", "position": { "x": 1.6, "y": 0.9, "z": 1.4 }, "rotation": { "x": 0, "y": 15, "z": 0 }, "scale": { "x": 0.1, "y": 0.2, "z": 0.1 }, "real": false, "ID": 3 }
] \\

Assistant Output: \\
Relevant Object(s):
[
  { "name": "Cup", "position": { "x": 1.6, "y": 0.9, "z": 1.4 }, "rotation": { "x": 0, "y": 15, "z": 0 }, "scale": { "x": 0.1, "y": 0.2, "z": 0.1 }, "real": false, "ID": 3 }
] \\

\#\# Example \\
User Input: \\ 
Position: (0, 0, 0) \\
Orientation: (0, 0, 0) \\
Target Objects: bush, table \\
Scene JSON file: 
[ 
  { "name": "Shrub", "position": { "x": 2, "y": 0, "z": 4 }, "rotation": { "x": 0, "y": 0, "z": 0 }, "scale": { "x": 1.0, "y": 1.0, "z": 1.0 }, "real": true, "ID": 1 },
  { "name": "Table", "position": { "x": 1.4, "y": 0.4, "z": 1.3 }, "rotation": { "x": 0, "y": 0, "z": 0 }, "scale": { "x": 1.2, "y": 0.7, "z": 1.9 }, "real": true, "ID": 2 }
] \\

Assistant Output: \\
Relevant Object(s): 
[ 
  { "name": "Table", "position": { "x": 1.4, "y": 0.4, "z": 1.3 }, "rotation": { "x": 0, "y": 0, "z": 0 }, "scale": { "x": 1.2, "y": 0.7, "z": 1.9 }, "real": true, "ID": 2 }
] \\

\#\# Example \\
Position: (1, 0, 2) \\
Orientation: (0, 90, 0) \\
Target Objects: chair, sofa \\
Scene JSON file: 
[
  { "name": "Chair", "position": { "x": 0.5, "y": 0, "z": 1.5 }, "rotation": { "x": 0, "y": 180, "z": 0 }, "scale": { "x": 0.8, "y": 1.0, "z": 0.8 }, "real": true, "ID": 1 },
  { "name": "Couch", "position": { "x": 2.0, "y": 0, "z": 1.0 }, "rotation": { "x": 0, "y": 0, "z": 0 }, "scale": { "x": 2.0, "y": 1.0, "z": 1.0 }, "real": false, "ID": 2 }
] \\

Assistant Output: \\
Relevant Object(s):
[
  { "name": "Couch", "position": { "x": 0.5, "y": 0, "z": 1.5 }, "rotation": { "x": 0, "y": 180, "z": 0 }, "scale": { "x": 0.8, "y": 1.0, "z": 0.8 }, "real": true, "ID": 1 },
  { "name": "Sofa", "position": { "x": 2.0, "y": 0, "z": 1.0 }, "rotation": { "x": 0, "y": 0, "z": 0 }, "scale": { "x": 2.0, "y": 1.0, "z": 1.0 }, "real": false, "ID": 2 }
]\\
\end{tcolorbox}
\subsection{Object Preparer}
\begin{tcolorbox}[colback=blue!5!white, colframe=blue!75!black, title=Object Preparer, breakable]
You will be given a string that has the name and the description of the object. Your task is to find out the dimensions of the object in the real world, determine its simple geometry shape (e.g., cylinder, cube, sphere, etc.), and decide if the object has gravity enabled.\\

\# Guidelines
- Output the size of the object in a common real-world size (x, y, z) dimension.
- Determine the basic geometry shape of the object (e.g., cylinder, cube, sphere) based on its general real-world appearance. \\
- If it’s a small object (for example, “car key”), slightly make the size larger for better visualization purposes.\\
- Put the dimension in the way that the object usually lies down in real life. For example, a lamp’s height will be the largest part, while a car key will have less height because it’s pretty flat while lying down.\\
- If the description specifies the scale of the object (for example, “large”), adjust the size you output accordingly.\\
- Decide if the object has gravity enabled:\\
  - For most physical objects, gravity should be `true`.\\
  - For floating objects (e.g., balloons) or non-physical objects, gravity should be `false`.\\
- If it doesn't seem like a physical object, return `0.5, 0.5, 0.5` for dimensions, `cube` as the shape, and `false` for gravity.\\

\# Examples

\#\# Example\\
User: \\
Object: chariity (green) \\ 
Description: large \\

Assistant:  \\
Dimensions: 0.5, 0.5, 0.5  \\
Shape: cube  \\
Gravity: false  \\

\#\# Example \\
User:  \\
Object: book (green)  \\
Description: large  \\

Assistant:  \\
Dimensions: 0.3, 0.1, 0.5  \\
Shape: cube  \\
Gravity: true  \\

\#\# Example \\
User:  \\
Object: table  \\
Description: none  \\

Assistant:  \\
Dimensions: 1.2, 0.8, 0.75  \\
Shape: cube  \\
Gravity: true  \\

\#\# Example\\
User:  \\
Object: lamp  \\
Description: tall\\  

Assistant:  \\
Dimensions: 0.15, 0.5, 0.15  \\
Shape: cylinder  \\
Gravity: true  \\

\#\# Example\\
User:  \\
Object: car key  \\
Description: none  \\

Assistant:  \\
Dimensions: 0.12, 0.04, 0.07  \\
Shape: cube  \\
Gravity: true  \\

\#\# Example\\
User:  \\
Object: coffee cup  \\
Description: medium  \\

Assistant:  \\
Dimensions: 0.1, 0.15, 0.1  \\
Shape: cylinder  \\
Gravity: true  \\

\#\# Example\\
User:  \\
Object: basketball  \\
Description: medium  \\

Assistant:  \\
Dimensions: 0.24, 0.24, 0.24  \\
Shape: sphere  \\
Gravity: true  \\
\end{tcolorbox}
\subsection{Location Optimizer}
\begin{tcolorbox}[colback=blue!5!white, colframe=blue!75!black, title=Location Optimizer, breakable]
You will be given a string containing an object's name, description, and its relationship to various locations. Your task is to determine the surface type where the object is placed or related. \\

\# Guidelines \\
- The possible surface types are: \\
  - **vertical**: A surface standing upright, like walls. \\
  - **on the top**: A flat horizontal surface where the object rests, like tables or shelves. \\
  - **hanging down**: A surface where the object is attached but hangs downward, like the underside of a table or a ceiling. \\
  - **floating**: If the object is not in direct contact with any surface or location (e.g., hovering or unsupported). \\
- For all the locations, identify the surface type based on the object's relationship to them. \\
- If multiple surface types are possible for a single location, group them in parentheses.\\
- If no valid surface type can be determined, output `unknown`.\\

\# Examples\\

\#\# Example 1\\
User:  \\
Object: poster (white, paper)  \\
Description: big  \\
Location 1: wall  \\
Relation to Location 1: on the surface  \\
Location 2: near  \\
Relation to Location 2: on  \\

Assistant:  \\
Surface Type: vertical  \\

\#\# Example 2\\
User:  \\
Object: chandelier (crystal, decorative)  \\
Description: luxurious  \\
Location 1: ceiling  \\
Relation to Location 1: hanging down  \\
Location 2: air  \\
Relation to Location 2: floating  \\

Assistant:  \\
Surface Type: hanging down  \\

\#\# Example 3\\
User:  \\
Object: book (blue, hardcover)  \\
Description: large  \\
Location 1: table  \\
Relation to Location 1: on  \\
Location 2: shelf  \\
Relation to Location 2: on  \\

Assistant:  \\
Surface Type: on the top  \\

\#\# Example 4\\
User:  \\
Object: curtain (red, fabric)  \\
Description: long  \\
Location 1: wall  \\
Relation to Location 1: near  \\
Location 2: rod  \\
Relation to Location 2: hanging down \\ 

Assistant:  \\
Surface Type 1: hanging down  \\

\#\# Example 5\\
User:  \\
Object: drone (metallic, quadcopter)  \\
Description: medium  \\
Location 2: table  \\
Relation to Location 2: on  \\

Assistant:  \\
Surface Type 2: on the top  \\

\end{tcolorbox}

\end{document}